\documentclass[aps,reprint,showpacs,preprintnumbers,amsmath,amssymb,prb,citeautoscript]{revtex4-1} 

\usepackage{bm}
\usepackage{color}
\usepackage{dcolumn}
\usepackage{graphicx}
\graphicspath{{figs/}{./}}
\newcommand\FigureFile[1] {#1.eps}
\DeclareGraphicsExtensions{pdf}

\newcommand\eq[1]                              
{
\begin{align*}
#1
\end{align*}
}

\newcommand\eql[2] 
{
\begin{equation}\label{#1}
\begin{split}
#2
\end{split}
\end{equation}
}

\newcommand\eqsl[1]                            
{
\begin{align}
#1
\end{align}
}

\newcommand\eqssl[2]                      
{
\begin{subequations}\label{#1}
\begin{align}
#2
\end{align}
\end{subequations}
}

\newcommand\Fig[1]     {Fig.~\ref{#1}}

\newcommand\Ref[1]     {Ref.~\onlinecite{#1}}

\newcommand\PsiT[1][]  {\Psi_{\mathrm{T}#1}^{}}

\newcommand\compPackage[1] {{\footnotesize{#1}}}

\newcommand\GAMESS     {\compPackage{GAMESS}}
\newcommand\NWCHEM     {\compPackage{NWCHEM}}

\newcommand\Ebind      {E_{b}}

\newcommand\Eh[1][]    {\ensuremath{E_\mathrm{h}#1}}

\newcommand\Ang        {\ensuremath{\textrm{\AA}}}

\newcommand\Order[1]   {\mathcal{O}\left(#1\right)}

\newcommand\Crtwo      {\textrm{Cr$_{\textrm{2}}$}}

\definecolor{xmgrace-green4}{rgb}{0.0,0.55,0.0}
\definecolor{Green}{rgb}{0.2,0.96,0.2}
\definecolor{Remarks}{rgb}{1,0.3,0.3}
\definecolor{Extra}{rgb}{0.2,0.2,1}
\definecolor{Blue}{rgb}{0.2,0.3,1}
\definecolor{Black}{rgb}{0,0,0}

\newcommand\COMMENTED[1] {}

\newcommand\Section[1]   {}
\newcommand\Subsection[1] {}
\newcommand\Subsubsection[1] {}

\begin{document}

\title{An Auxiliary-Field Quantum Monte Carlo Study of the Chromium Dimer}

\author{Wirawan Purwanto}
\email{wirawan0@gmail.com}
\affiliation{Department of Physics, College of William and Mary,
Williamsburg, Virginia 23187-8795, USA}

\author{Shiwei Zhang}
\affiliation{Department of Physics, College of William and Mary,
Williamsburg, Virginia 23187-8795, USA}

\author{Henry Krakauer}
\affiliation{Department of Physics, College of William and Mary,
Williamsburg, Virginia 23187-8795, USA}

\date{\today}

\begin{abstract}

The chromium dimer ({\Crtwo}) presents an outstanding challenge for
many-body electronic structure methods.
Its complicated
nature of binding, with a formal sextuple bond and an unusual potential energy curve,
is emblematic of the competing tendencies and delicate balance found in many strongly correlated materials.
We present a near-exact calculation of
the potential energy curve (PEC)
and ground state properties of {\Crtwo},
using the auxiliary-field quantum Monte Carlo (AFQMC) method.
Unconstrained, exact AFQMC calculations are first carried out for
a medium-sized
but realistic basis set.
Elimination of the remaining finite-basis errors and extrapolation to the complete basis set  (CBS) limit is then achieved with
a combination of phaseless and
exact AFQMC calculations.
Final results for the PEC and spectroscopic constants are in excellent agreement with experiment.

\end{abstract}


\maketitle

\Section{Introduction}

The chromium dimer
is a strongly correlated molecule which
poses a formidable challenge to
even the most accurate
many-body methods.
It features a formal sextuple bond, with a weak binding energy
($\sim 1.5$\,eV), a
short equilibrium bond length
($\sim$ 1.7\,\Ang),
and an unusual ``shoulder'' structure in its potential energy curve (PEC).\cite{Bondybey1983,Simard1998,Casey1993}
The ground state of {\Crtwo}
is highly multiconfigurational, and
proper theoretical description
requires an accurate treatment
of the strong $3d$ electron correlations (both static and dynamic).
The nature of the PEC in {\Crtwo}
is representative of the competing tendencies separated by small energy differences seen in many
strongly correlated materials.
Because of the fundamental and technological significance of such materials,
improving our abilities for accurate calculations in strongly correlated systems is one of the most
pressing needs in condensed matter physics and quantum chemistry.

\begin{figure}[!htbp]
\includegraphics[scale=0.35]{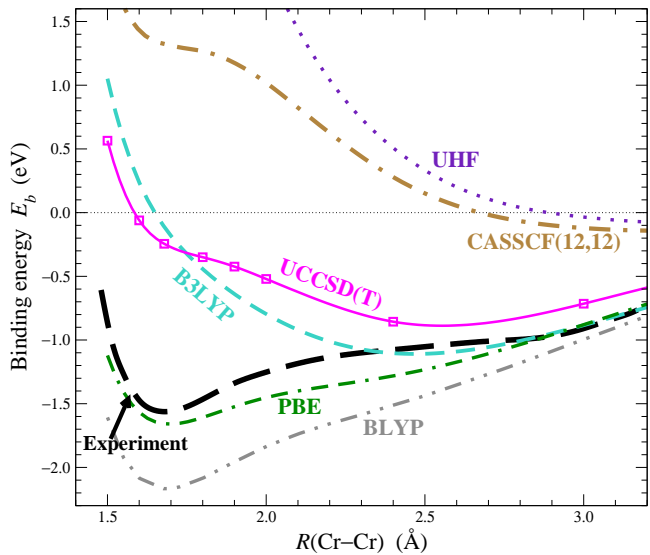}
\caption{\label{fig:Cr2-PEC-std-QC}
(Color online)
Representative results of the {\Crtwo} PEC
from standard quantum chemistry calculations.
Shown are results from unrestricted HF (UHF),
complete active space self-consistent field (CASSCF) with
12 active electrons and 12 active orbitals,
UCCSD(T) with UHF reference wave function,
and DFT
with various exchange-correlation functionals
(B3LYP,\cite{Becke1993,Stephens1994} PBE,\cite{Perdew1996} BLYP\cite{Becke1988,Lee1988}).
Most calculations use the cc-pwCVTZ-DK basis set,
except UCCSD(T), where extrapolation to
the CBS
limit was done, as described later in the text.
Experimental PEC was taken from \Ref{Casey1993}.
}
\end{figure}

Standard quantum chemistry methods, such as density functional theory (DFT),
Hartree-Fock (HF), and post-HF methods such as single-reference
second-order M{\o}ller-Plesset perturbation theory (MP2) and
single-reference coupled cluster with singles, doubles, and perturbative
triples [CCSD(T)], all fail to describe the correct binding of {\Crtwo}.
Representative standard quantum chemistry results are
shown in \Fig{fig:Cr2-PEC-std-QC}.
As often is the case, the DFT results vary greatly,
depending on the choice of
exchange-correlation functional.
There have also been numerous attempts to calculate the PEC of {\Crtwo}
using sophisticated multireference quantum chemistry methods
\cite{Stoll1996,Dachsel1999,Angeli2002,Angeli2006,Celani2004,Muller2009},
including the complete active space second-order perturbation theory
(CASPT2) \cite{Andersson1995,Roos1995,Roos2003}
and, more recently, CASPT2 based on a large
density matrix renormalization group (DMRG)
reference wave function (DMRG-CASPT2).\cite{Kurashige2011}
These calculations
obtain qualitatively correct binding, but the results are sensitive to
choice of active space and/or basis set.
Standard quantum Monte Carlo (QMC) approaches \cite{Need2010,Kolorenc2011} have also been severely challenged.
A recent fixed-node diffusion Monte Carlo (DMC) study, which examined the use of
a variety of single- and multi-determinant trial wave functions, did not obtain satisfactory
binding (indeed the molecular energy was found to be higher than
the sum of two isolated atoms). \cite{Hongo2012}
All these underline the extreme challenge in achieving
an accurate
theoretical description of the {\Crtwo} PEC.

In this paper we present
calculations of the {\Crtwo} PEC and
ground-state properties using the
auxiliary-field quantum Monte Carlo (AFQMC) method.
\cite{Zhang2003,AlSaidi2006b,Suewattana2007,Zhang2013}
We first describe
exact free-projection (FP) AFQMC calculations,
where we release the usual phaseless approximation, \cite{Zhang2003} which is used to control the phase/sign problem \cite{Zhang2013}.
The results are then extrapolated to the
CBS limit using
a combination of phaseless and exact AFQMC calculations.
Final results for the PEC and spectroscopic constants are in excellent agreement with experiment.

AFQMC obtains ground-state properties
by stochastically sampling the many-body ground-state wave function in the space
of Slater determinants, expressed in a chosen one-particle basis.
\cite{Zhang2003,AlSaidi2006b,Suewattana2007,Zhang2013}
It has modest polynomial scaling with system size $M$
[$\Order{M^3}$ or $\Order{M^4}$]
rather than the exponential scaling of CI calculations, or
the high-order polynomial scaling of typical quantum chemistry many-body methods.
The
FP AFQMC  \cite{Purwanto2009_Si,Shi2013}, which leaves the fermion sign/phase problem uncontrolled \cite{Zhang2013}, is exact but has exponential scaling due
to rapidly increasing stochastic noise with projection imaginary-time.
The AFQMC phaseless approximation  (ph-AFQMC) \cite{Zhang2003} was introduced to control this,
resulting in a practical method which restores the low computational scaling. The method
has been demonstrated to yield accurate results in
many atomic, molecular, cluster, and extended systems.
\cite{Zhang2003,AlSaidi2006b,Suewattana2007,Purwanto2009_C2,Virgus2014}
For {\Crtwo},
we have found that the current implementation of 
the phaseless approximation (using standard single- or multi-determinant trial wave functions),
while leading to a qualitatively correct PEC, exhibits noticeable systematic error in the binding energy.
To eliminate the residual systematic errors,
we are able to carry out exact,
large-scale FP-AFQMC calculations
using a moderate-sized but realistic basis set.
These exact results are used to benchmark
other many-body methods, including ph-AFQMC as well as previously published results.
In a final step, the results are combined with ph-AFQMC calculations with large basis sets
to obtain a near-exact PEC in the CBS limit.

\Section{Realistic basis AFQMC calculations}
\label{sec:Cr2-real}

The AFQMC calculations reported here employ
standard quantum chemistry gaussian type orbital
basis sets. \cite{EMSL_BasisSets2007}
Our calculations employed the Douglas--Kroll--Hess
scalar-relativistic all-electron Hamiltonian, with core-valence
correlation-consistent Gaussian basis sets,
cc-pwCV$x$Z-DK, with $x = 3, 4, 5$.
(We will hereafter refer to these as TZ, QZ, and 5Z, respectively.)
For a chosen basis, AFQMC thus
treats
the same Hamiltonian
as that of a corresponding
many-body quantum chemistry calculation, allowing, for example, direct
comparisons of absolute total energies.
This was done below
with DMRG calculations, where results
using a
small split-valence (SV) basis
\cite{Schaefer1992} were available \cite{Kurashige2009}. 
AFQMC projects the ground state starting from a trial wave function $\PsiT$, which is also used
in the mixed-estimator to compute the ground-state energy and additionally
in ph-AFQMC to control the fermionic sign/phase problem.
We used two choices of $\PsiT$, the
unrestricted Hartree-Fock (UHF) and truncated
complete active space self-consistent field (t-CASSCF)
as in our earlier work. \cite{Purwanto2009_C2}
CASSCF(12,12) was used, which
fully correlates 12 active electrons in
12 orbitals derived from the $3d$ and $4s$ atomic states.
The CASSCF wave function is truncated such that the weight
(squared coefficient) of the retained
determinants is $\sim 90\% - 92\%$ of the total.
This particular way of choosing the t-CASSCF  $\PsiT$
becomes increasingly 
expensive as the atoms are stretched from the equilibrium
bond length;
for larger bond lengths, 
broken spin symmetry UHF $\PsiT$
were used for FP-AFQMC, as discussed below.
All calculations used the frozen-core approximation, \cite{Purwanto2013}
freezing neon-core orbitals calculated at a lower level of theory (HF here).
The frozen-core Hamiltonian one- and two-body matrix elements and $\PsiT$ were
obtained
using outputs from modified quantum chemistry codes, {\NWCHEM}\cite{NWChem-6.0} and
{\GAMESS}\cite{Gamess}. 

\Subsection{Results for TZ basis}

\begin{figure}[t]
\includegraphics[scale=0.33]{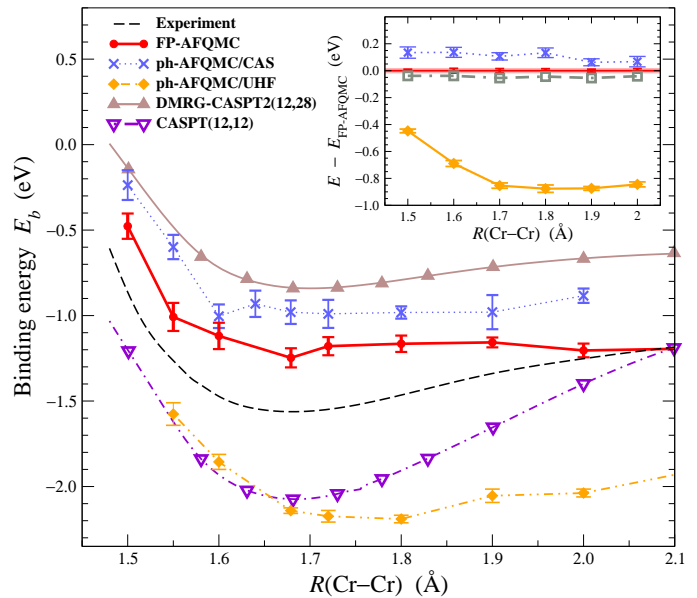}
\caption{\label{fig:Cr2-TZ-PEC}
(Color online)
Benchmarking the {\Crtwo} PECs in finite basis sets.
The main figure shows the calculated binding energy with the cc-pwCVTZ-DK basis, comparing
exact FP-AFQMC PECs to ph-AFQMC, 
DMRG-CASPT2 and CASPT2 (see text) from \Ref{Kurashige2011}.
The experimental PEC is also shown for reference.
The inset shows discrepancies of  \emph{total energies} compared to exact FP-AFQMC for DMRG (\Ref{Kurashige2009}) 
and ph-AFQMC for the small SV basis. \cite{Schaefer1992}
DMRG results are shown as empty squares, while AFQMC results have the same symbols as in 
the main figure.
}
\end{figure}

Figure~\ref{fig:Cr2-TZ-PEC}  presents benchmark results of the calculated {\Crtwo}  PECs.
The main figure compares AFQMC to CASPT2 calculations using the realistic  TZ basis set.
The inset compares AFQMC to DMRG calculations for the small SV basis.
The binding energy is given by $\Ebind \equiv E_{\textrm{mol}} - 2E_{\textrm{atom}}$, where the molecular and
atomic energies are calculated with the same method.
The FP-AFQMC calculations are exact for the chosen basis.
(While we have included the experimental PEC in the figure for reference, 
it should not
be directly compared to FP-AFQMC using the TZ basis. 
Extrapolation
to the CBS limit is discussed below.)
These calculations were done with $\sim 10^5$ or more walkers,
with an imaginary-time projection of 400-500 steps and a time-step of $0.02\,\Eh^{-1}$,
requiring significant computing resources.
We have verified that the
Trotter error in the calculated FP-AFQMC binding energy is smaller than the statistical error bar.
Phaseless AFQMC results are also shown, using UHF and t-CASSCF $\PsiT$, 
denoted ph-AFQMC/UHF and ph-AFQMC/CAS, respectively.
These
were done with $\sim 2000$ walkers,  
and are essentially in the 
zero Trotter time-step limit. 
The  ph-AFQMC/CAS PEC shows a small non-parallelity error (NPE) 
and is $\sim 0.2$~eV above the exact result. 
The ph-AFQMC/UHF PEC shows larger discrepancies, lying below FP-AFQMC by $\sim 0.6 - 1.0$~eV.
(Since ph-AFQMC is non-variational, \cite{Carlson1999}
this is possible, especially with a poor $\PsiT$.)
These AFQMC results are compared to two calculations using
CASPT2.
The PEC labeled CASPT2 is based on
a CASSCF(12,12) reference wave function, while DMRG-CASPT2 labels the PEC
based on a reference wave function from DMRG.
\cite{Kurashige2011}
The DMRG reference 
wave function is for an active space of 12 electrons and
28 orbitals,
which is much larger than that of the CASSCF(12,12).
Both CASPT2 and DMRG-CASPT2 differ from the exact PEC:
CASPT2 overbinds by $\sim 1$~eV, while the nominally improved DMRG-CASPT2
is underbound. 
This comparison shows that even with the larger DMRG active space,
the
CASPT2 treatment still does not
fully account for
dynamical correlation.
\Subsection{Minimal basis benchmark calculations}
\label{sec:Cr2-bench}
%
%
In contrast, the full DMRG result,
available for the small SV basis,
agrees with exact FP-AFQMC to
within the stochastic error of the latter, as seen in the inset of \Fig{fig:Cr2-TZ-PEC}.
(The SV basis benchmark calculations were done with
a non-relativistic Hamiltonian with 12 core electrons frozen.
With this small basis, the molecule is not bound.)
Similar to the TZ-basis results in the main figure, ph-AFQMC/CAS lies
above the exact FP-AFQMC curve by $\lesssim 0.15$~eV with a small
NPE $\lesssim 0.08$~eV,
while ph-AFQMC/UHF shows larger errors.


The benchmark results in \Fig{fig:Cr2-TZ-PEC}  illustrate
the importance of $\PsiT$ for ph-AFQMC
in the strongly correlated {\Crtwo}.
Although the UHF and  t-CASSCF wave functions have similar variational energies,
t-CASSCF is a
better $\PsiT$ because
it more accurately describes the multiconfigurational
nature of the ground state and, unlike UHF,
does not break spin symmetry (to within small truncation error).
For example, at stretched geometries $R \gtrsim 1.9\,\Ang$,
UHF has a lower variational energy than
the t-CASSCF $\PsiT$ at $90\%$ total weight; but the UHF $\PsiT$ has large
spin-contamination 
${S}^2  \gtrsim  5$.
This leads to significant errors in ph-AFQMC/UHF
and long imaginary-time equilibration times. \cite{Purwanto2008}
The AFQMC/CAS approach, on the other hand, becomes increasingly
expensive as the atoms are stretched from the equilibrium
geometry, because the number of the required determinants in $\PsiT$ grows rapidly.
(At $2.0\,\Ang$, for example, a $92\%$ cut retains
$\sim 1800$ determinants.)
For FP-AFQMC, which is not biased by $\PsiT$,
calculations at larger bond lengths were carried out by
initializing an approximately spin-pure walker population using an
aggressively truncated t-CASSCF wave function,
while the energy mixed-estimator was evaluated using the UHF $\PsiT$. 
This approach reduces the time to equilibrate FP-AFQMC.

\Subsection{CBS extrapolation}

\begin{figure}[!hbtp]
\includegraphics[scale=0.305]{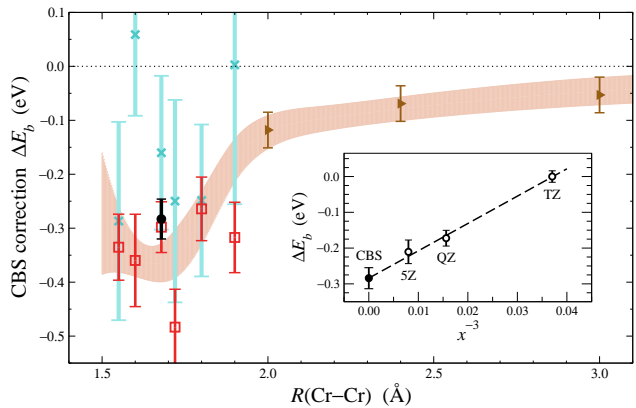}
\caption{\label{fig:Cr2-CBX}
(Color online) CBS correction to the correlation contribution
of the binding energy, with respect to cc-pwCV$x$Z-DK, as a function of
Cr--Cr distance.
The final CBS correction is shown by the shaded band in the main figure, where
the shading width represents combined stochastic and fitting errors. 
The red open squares are from ph-AFQMC/UHF, blue crosses from ph-AFQMC/CAS, while the 
brown triangles are from free projection.
The inset verifies the accuracy of the TZ--QZ extrapolation,
using a ph-AFQMC/UHF 5Z calculation at $1.68$\,{\AA} (the black filled circle in the main figure), where
good linearity is seen in the plot of  cc-pwCV$x$Z-DK results 
vs.~the basis set cardinal number $x$.
}
\end{figure}


The final CBS correction, 
which is to be added to the cc-pwCVTZ-DK FP-AFQMC results in Fig.~\ref{fig:Cr2-TZ-PEC},
consists of a (small) HF contribution plus a correlation contribution
$\Delta E_b$.\cite{Purwanto2011}
The latter
is given by the shaded curve in \Fig{fig:Cr2-CBX}.
It was obtained by extrapolating TZ and QZ results, with cross-check from 5Z calculations, as follows.
For $R < 2.0$\,{\AA}, we performed ph-AFQMC calculations using both UHF and t-CASSCF trial 
wave functions to extrapolate to the CBS limit \cite{Helgaker1997,Purwanto2011}. 
Although the 
ph-AFQMC/UHF PEC lies below the exact result, while ph-AFQMC/CAS lies above,
their respective $\Delta E_b$ are in good agreement. 
The UHF results, which have considerably smaller statistical error bars, are used to obtain the smooth fit.
We also performed a ph-AFQMC/UHF 5Z calculation at $R = 1.68$\,{\AA}
to check the accuracy of the TZ--QZ extrapolation, as shown in the inset.
For $R \ge 2.0$\,{\AA}, we used an alternative approach to obtain $\Delta E_b$,
since both ph-AFQMC/UHF and ph-AFQMC/CAS have some difficulties
in this region, as discussed earlier.
FP-AFQMC/UHF TZ and QZ calculations were performed for a larger frozen core, 
which also freezes the semicore $3s$ and $3p$ orbitals.
For the larger $R$,
the neglect of semicore correlation effects has negligible effect on
 $\Delta E_b$, as we confirmed 
with UCCSD(T) TZ and QZ calculations. 
With fewer correlated electrons, stochastic errors were reduced,
allowing us to extract  $\Delta E_b$ with FP-AFQMC.
(We also found that $\Delta E_b$ converged well before full equilibration with both the TZ and QZ basis.)
A smooth fit was made to these results for $R\ge 2.0$\,{\AA},
with a spline joining the two regions to yield the final  $\Delta E_b$ for the entire PEC.

\begin{figure}[!htbp]
\includegraphics[scale=0.33]{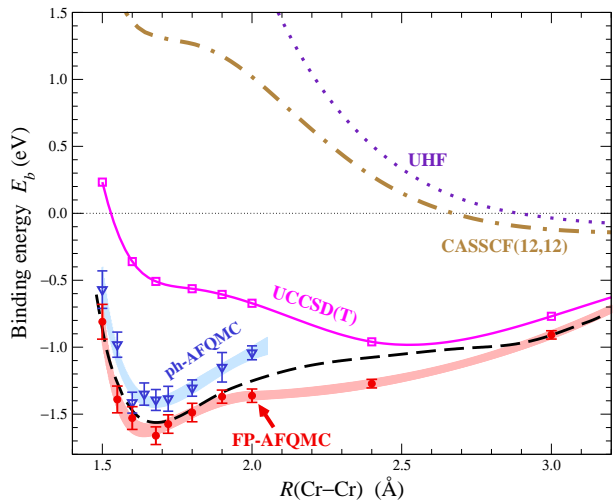}
\caption{\label{fig:Cr2-final-cbs}
(Color online)
CBS extrapolated ph-AFQMC and exact FP-AFQMC PECs compared to experiment (dashed black line).
Results from several standard quantum chemistry methods are also shown for reference.
}
\end{figure}

The CBS-extrapolated FP-AFQMC PEC, shown in \Fig{fig:Cr2-final-cbs},
is generally in excellent agreement with experiment,
except for the shoulder region, which is discussed further below.
The corresponding spectroscopic constants, obtained from both FP- and ph-AFQMC, are compared
to experiment in Table~\ref{tbl:Cr2-consts}.
The UHF, CASSCF(12,12), and UCCSD(T) PECs (reproduced from  \Fig{fig:Cr2-PEC-std-QC})
do not show binding [although UCCSD(T) has an outer well near 2.7\,{\AA}]. 
Both CASSCF(12,12) and UCCSD(T) evidence a plateau-like feature
at short Cr-Cr distance, however. 
As previously seen in other applications, 
the ph-AFQMC recovers from a qualitatively incorrect $\PsiT$ from UHF or CASSCF(12,12).
Although the ph-AFQMC/CAS is somewhat underbound,
the equilibrium bond length and vibrational frequency agree
very well with experiment as shown in Table~\ref{tbl:Cr2-consts}.
The FP-AFQMC result is seen to lie somewhat lower than experiment
in the
shoulder region
\mbox{$\simeq 2.0$ -- 2.7 \AA},
which is where the experimental PEC~\cite{Casey1993} has the greatest
uncertainty.
The experimental PEC was based on high-resolution photoelectron spectra of Cr$_2^{\,-}$,
which showed 29 vibrationally resolved transitions to the neutral Cr$_2$ ground state.
As noted in Ref.~\onlinecite{Casey1993}, there were large
gaps in the vibrational data between 3040 and 4880 cm$^{-1}$, which
insufficiently constrained the shape of the potential in this
region.
\begin{table}[htbp]
\caption{\label{tbl:Cr2-consts}
Spectroscopic constants of {\Crtwo} computed using phaseless and
free-projection AFQMC methods, extrapolated to the
CBS limit.
$E_b$ is the molecular binding energy
(zero-point energy has been removed from the experimental value);
$R_0$ is the equilibrium bond length;
and $\omega_e$ is the harmonic vibrational frequency.
}
{%
\newcommand\Xa{\footnote{\Ref{Simard1998}}}
\newcommand\Xb{\footnote{\Ref{Bondybey1983}}}
\newcommand\Xc{\footnote{\Ref{Casey1993}}}
\newcommand\Xd{\footnote{\Ref{Hilpert1987}}}
\begin{ruledtabular}
\begin{tabular}{lclll}
Method            &          & $E_b$ (eV)   & $R_0$ (\AA)  & $\omega_e$ (cm$^{-1}$) \\
\hline
ph-AFQMC          &          & $-1.42(4)$   & $1.68(2)$   & $520(59)$  \\
FP-AFQMC          &          & $-1.63(5)$   & $1.65(2)$   & $552(93)$  \\
\hline
Experiment
                  &          & $-1.56(6)$\Xa& $1.6788$\Xb & $480.6(5)$\Xc\\
                  &          & $-1.47(5)$\Xd&             &             \\ 
\end{tabular}
\end{ruledtabular}
}
\end{table}
The possibility was stated that the true PEC could actually have
a shallow minimum where the experimentally fitted PEC exhibits a shoulder.
Future theoretical study, with reduced stochastic uncertainty and at more bondlengths, is warranted to further
assess the shape of the PEC in this region.

\Section{Summary}
\label{sec:Summary}

In summary, we have presented 
a near-exact calculation of
the PEC
and spectroscopic properties of {\Crtwo},
using the
AFQMC method.
Unconstrained, exact AFQMC calculations were first carried out for
a medium-sized
but realistic basis set.
Elimination of the remaining finite-basis errors and extrapolation to the CBS limit was then achieved with
a combination of phaseless and
FP AFQMC calculations.
This hybrid approach enabled us to obtain the most accurate
theoretical results of {\Crtwo} ground-state properties
obtained to date,
which are in excellent agreement with experiment.

\begin{acknowledgments}

This work was supported by
DOE (DE-FG02-09ER16046),
NSF (DMR-1409510),
and
ONR (N000141211042).
We acknowledge a DOE CMCSN grant (DE-FG02-11ER16257)
for facilitating stimulating interactions.
An award of computer time was provided by
the Innovative and Novel Computational Impact on Theory and Experiment
(INCITE) program,
using resources of the Oak Ridge Leadership Computing Facility (Titan)
at the Oak Ridge National Laboratory,
which is supported by the Office of Science of the U.S. Department of Energy
under Contract No. DE-AC05-00OR22725.
Some of the earlier computing was performed on Blue Waters, which is supported by the National Science Foundation (award
number OCI 07-25070) and the state of Illinois. Blue Waters is a joint
effort of the University of Illinois at Urbana-Champaign and its National
Center for Supercomputing Applications.
We also acknowledge the computing support from the Center for Piezoelectrics by Design.
The authors would like to thank Yudistira Virgus, Eric J. Walter, and
Simone Chiesa for
many useful discussions.

\end{acknowledgments}

\bibliography{AFQMC-bib-entries}

\end{document}